\documentstyle[eqsecnum,prd,aps,epsfig]{revtex}
\begin{document}
\draft
\title{Reduction of neutrino - nucleon scattering rate by nucleon -
nucleon collisions}
\author{Shoichi Yamada 
\thanks{e-mail : shoichi@phys.s.u-tokyo.ac.jp, 
TEL : +81-3-5841-4191, 
FAX : +81-3-5841-4224}
}
\address{Department of Physics, School of Science, University of
Tokyo\\
7-3-1 Hongo, Bunkyo-ku, Tokyo 113-0033, JAPAN}
\date{\today}
\maketitle
\begin{abstract}
We studied possible modifications of the neutrino -- nucleon
scattering rate due to the nucleon -- nucleon collisions in the
hot dense matter which we find in the supernova core. We show that the finite
width of the nucleon spectral function induced by the nucleon collisions
leads to broadening of the dynamical spin structure function 
of the nucleon, resulting in the reduction of the rate of neutrino --
nucleon scattering via the axial vector current and making the energy
exchange between neutrinos and nucleons easier. The reduction rate is 
relatively large $\sim$0.6 even at density $\sim 10^{13}$g/cm$^{3}$ and 
could have a significant impact on the dynamics of the collapse-driven 
supernova as well as the cooling of the proto neutron star. 
\end{abstract}

\section{introduction}
The collapse-driven supernova is supposed to be an outcome of
gravitational collapse of a massive star $\gtrsim 8 {\rm M}_{\odot}$
at the end of its evolution. Although more than 60 years have passed
since the original idea by Baade and Zwicky\cite{bz34a,bz34b}, and the detection of
neutrinos from SN1987A\cite{imb87,kam87} confirmed that 
our scenario is correct on the
whole, we have not yet figured out how this phenomenon occurs. 

These days many researchers of the supernova consider that neutrinos diffusing 
out of the proto neutron star play a crucial role in heating the
material behind the shock wave and expelling the outer layer of the
star\cite{wil82,bw85}. It turned out, however, that this mechanism is 
quite sensitive to the neutrino luminosity\cite{bg93,jm93}. In fact, 
Janka and M\"{u}ller\cite{jm93} showed by
numerical simulations that only about 20\% of increase in neutrino
luminosity could lead to a successful explosion of a model which otherwise
failed to explode. Hence mechanisms to boost the neutrino luminosity
have been quested. Although many authors have been devoted in the
study of convection in the core and it has been shown that the
convection does help neutrinos heat matter, it is still controversial
if the convection alone is sufficient for successful explosion or
not\cite{jm93,jm96,he94,bu95,sh94}. 
It is also found that the sophistication of the numerical 
treatment of neutrino transport could increase the neutrino
luminosity although the quantitative assessment of the
difference it makes in the realistic context remains to be 
done\cite{mmbg98,yjs99,bu99}. 

On the other hand, our knowledge of the neutrino reaction rates in the 
hot dense medium is rather poor. In fact, even in the most elaborate
simulations of supernovae it has been assumed that the reactions of
neutrinos with nucleons are the same as in vacuum and the effect 
of surrounding matter is ignored\cite{br85,mb93a,mb93b,mb93c}. 
However, the  wavelength of a 
30MeV neutrino, for example, is longer than the mean 
separation of nucleons for the density 
$\gtrsim 10^{13}$g/cm$^{3}$, and the time scale corresponding to the same
energy is roughly of the same order as the mean free time of nucleon 
between collisions. Hence we have to study the spatial and temporal
correlations of the matter, and it could be possible that the
many body effects change the opacity for neutrinos considerably and we
get the desired enhancement of neutrino luminosity and/or energy. In fact, 
efforts to find a possible modification of rates of neutrino reactions 
with nucleons have been made 
by several 
authors\cite{iw82,sa89,hw91,sw95,rs95,rs96,rs98,bs98,bs99,re98b,re98c,yt99}. 
In these studies
neutrino - nucleon scatterings are one of the targets, since it is one
of the major sources of opacity for neutrinos. 
Reddy et al.\cite{re98b,re98c} pointed
out that taking a correct effective mass of nucleon into account
changes the scattering rate in the high density regime $\gtrsim \rho
_{0}$ which is the saturation density. Horowitz et al.\cite{hw91}, 
Burrows et al.\cite{bs98,bs99}, Reddy et al.\cite{re98b,re98c} and 
Yamada et al.\cite{yt99} discussed the correlation effects 
due to the particle - hole excitation using a random phase approximation 
(RPA). In these studies the nucleons were assumed to be quasi-particles 
with the vanishing width of the spectral functions. On the other
hand, Raffelt and his collaborators\cite{rs95,rs96,rs98} claimed that the neutrino
scattering rate could be also reduced by losing the temporal
correlations of the spins of nucleon, thus broadening the width of the 
structure function due to collisions of the nucleon with other nucleons surrounding
it (see also \cite{sw95}). This is a counter part for scattering of the so-called 
Landau-Pomeranchuk-Migdal effect\cite{lp53a,lp53b,mg56} 
for the bremsstrahlung. It was also
pointed out that this broadening of the structure function 
could enhance the inelasticity of the scattering and affect the 
energy spectra of neutrinos\cite{hr98}, since the width of the
structure function is of the order of the matter temperature at a 
relatively low density $\sim 10^{13}$g/cm$^{3}$ according to their 
estimate. In these studies they assumed that the structure 
function of nucleon takes the Lorentzian form and inferred its width from the 
typical collision rate of nucleon with the overall normalization 
determined by the sum rule and the detailed balance imposed by
hand. In addition, they only discussed long wave length limits. 
In this paper we study this collisional effect on 
the neutrino -- nucleon scattering rates on the field theoretical 
basis. The width of the spectral function of nucleon is evaluated by
solving the Bethe-Salpeter equation in medium and calculating the
imaginary part of the nucleon self-energy\cite{srs90,ar96,rs99}, 
and then it is used to
calculate the structure functions of nucleon as a function of the transfered 
four momentum\cite{kv95}. In so doing the separable 
Yamaguchi potential\cite{ym54} was used for simplicity. 

The paper is organized as follows. In the next section we formulate the 
basic equations. The results are presented in section 3. Discussion is 
given in the last section.
    
\section{formulation}

\subsection{scattering rates}

The scattering rates of neutrinos with nucleons are quite generally 
formulated as : 
\begin{eqnarray}
\label{eq:decom}
\nonumber
R\,(E_{\nu }^{in},\, E_{\nu }^{out}, \, \cos \theta) \  = \ 4 \, G_{F}^{2} \,
E_{\nu }^{in} E_{\nu }^{out} \ [ R_{1}(k) \, (1 + \cos \theta) \ 
& + & \ R_{2}(k) \, (3 - \cos \theta) \\ 
& - & \ 2 \, (E_{\nu }^{in} + E_{\nu }^{out}) \, R_{5}(k) \, 
(1 - \cos \theta)] \quad ,
\end{eqnarray}
where $E_{\nu }^{in}$ and $E_{\nu }^{out}$ are the incident and
outgoing neutrino energy, respectively, $\theta $ is the
scattering angle, and $k$ is the four momentum transferred from
neutrino to nucleon. In the right hand side of 
Eq.~(\ref{eq:decom}), the third term is usually much smaller than 
the other two, so that we ignore it in the following. The so-called dynamical 
structure functions $R_{1}$ and $R_{2}$ are given in the 
non-relativistic limit which we assume in the following, as
\begin{eqnarray} 
R_{1}(k) & \approx & \ h_{V}^{2} \, \int \! d^{4} \! x \ e^{ikx} \ 
\langle \, \rho_{N}(x) \, \rho_{N}(0) \, \rangle \quad , \\
\label{eq:r2}
R_{2}(k) & \approx & \ \frac{h_{A}^{2}}{3} \, \int \! d^{4} \! x \ e^{ikx} \ 
\langle \, \bbox{s}_{N}^{i}(x) \, \bbox{s}_{N}^{i}(0) \, \rangle
\quad .
\end{eqnarray}
Here $h_{V}$ and $h_{A}$ are the weak coupling constants of 
nucleons, $\langle \, \cdots \, \rangle$ stands for the thermal ensemble average 
of the argument, and $\rho_{N}(x)$ and $\bbox{s}_{N}^{i}(x)$ are the density
and the spin density of nucleon. Thus it is obvious that $R_{1}(k)$ 
and $R_{2}(k)$ are nothing but their correlations in the
matter. As is clear from the factor of the $R_{2}(k)$ in
Eq.~(\ref{eq:decom}), $R_{2}(k)$ is more important than
$R_{1}(k)$. Thus we will discuss in this paper the modification of 
$R_{2}(k)$ , or the spin-density correlation function, due to the 
nucleon -- nucleon collisions in the supernova core. 

\subsection{temporal spin-density correlation of a nucleon in random walk}

\begin{figure}
\begin{center}
\epsfig{file=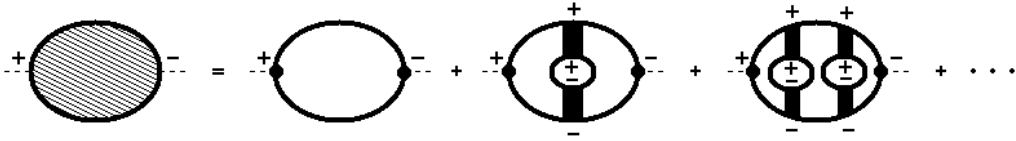,width=.80\textwidth}
\caption{The Feynman diagrams contributing to the spin density
correlation function. \label{fig1}}
\end{center}
\end{figure}

Just by the same argument as made by Knoll et al.\cite{kv95} for the momentum
correlation, the spin-density correlation of the nucleon 
which is successively scattered by other nucleons with a mean
collision rate, $\Gamma $, changing the
direction of its spin via spin-dependent interactions, is given by 
ignoring the spatial non-uniformity as
\begin{equation}
S^{ik}(\tau ) \ = \ \langle s^{i}(\tau ) \, s^{k}(0) \rangle \ = \  
{\rm e}^{-|\Gamma \tau |} \sum _{n = 0}^{\infty } 
\frac{|\Gamma \tau |^{n}}{n !} \, \langle s_{m}^{i} \, s_{m+n}^{k}
\rangle _{m} \quad .
\end{equation}
In the above equation, $n$ is the number of the scatterings during the 
time interval of $\tau $. $\bbox{s}_{m}$ is the spin-density at the m-th
collision, and $\langle \cdots \rangle _{m}$ means taking the average
of the argument with respect to the ensemble of the sequence of $n$
scatterings. Taking the Fourier transform of $S^{ik}(\tau )$, we
obtain 
\begin{equation}
\label{eq:sik}
S^{ik} (\omega ) \ = \ \sum _{n = 0}^{\infty } \, 
\langle s_{m}^{i} \, s_{m+n}^{k} \rangle _{m} \, 
\frac{2 \Gamma ^{n} Re(\Gamma + i \omega )^{n + 1}}{(\omega ^{2} +
\Gamma ^{2})^{n + 1}}  \quad .
\end{equation}    
It was pointed out in the paper by Knoll et al.\cite{kv95} that the above
summation of the Lorentz functions correspond to the summation of the
Feynman diagrams depicted in Fig.~\ref{fig1} in the closed time path
formalism. 

\begin{figure}
\begin{center}
\epsfig{file=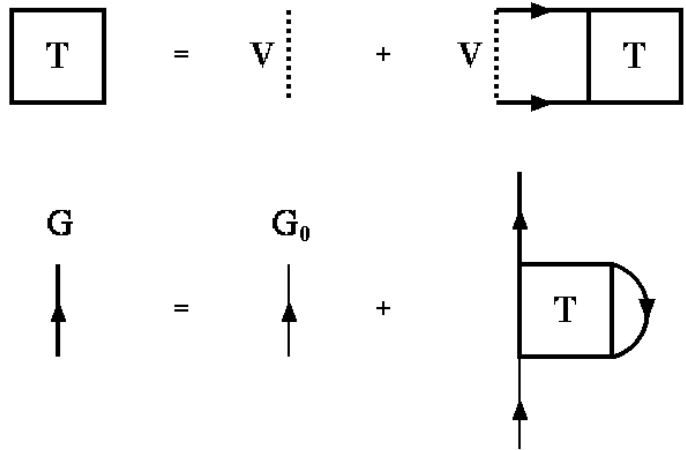,width=.60\textwidth}
\caption{The Bethe-Salpeter equation of the $T-$matrix for two
nucleons (upper) and the Dyson equation for the Green function of
nucleon (lower). \label{fig2}}
\end{center}
\end{figure}

If we further assume that the spin-density auto-correlation
decreases by a constant rate, $\alpha $, after each scattering, 
$\langle s_{m}^{i} \, s_{m+n}^{k} \rangle _{m} = \alpha ^{n}
\langle s_{m}^{i} \, s_{m}^{k} \rangle _{m}$, then we obtain 
\begin{eqnarray}
S^{ik} (\omega ) & = & \frac{2 \, \Gamma '}{(\omega ^{2} + {\Gamma '}^{2})}
\, \langle s_{m}^{i} \, s_{m}^{k} \rangle _{m} \\
\Gamma ' & = & (1 - \alpha ) \, \Gamma  \quad .
\end{eqnarray}  
This implies that the higher order terms account for the difference
between the collision rate, $\Gamma $, and the relaxation time of
spin-density, $\Gamma '$. If the spin flip is fast enough, $\alpha
\approx 0$, these two rates become almost identical. In this
case, the spin correlation function of Eq.~(\ref{eq:r2a}) is
approximated by the first diagram in Fig.~\ref{fig1}. Note that the
propagator in that diagram is not a free propagator but a dressed one
in the medium. The
self-energy included in the dressed propagator comes from the nucleon
collisions with other nucleons. In this paper this self-energy is 
evaluated by solving the Bethe-Salpeter equation for two nucleons
as drawn schematically in Fig.~\ref{fig2}\cite{srs90,ar96,rs99}. 
In so doing, we assumed a
separable Yamaguchi potential\cite{ym54} in order to facilitate solving the 
Bethe-Salpeter equation. Although this is an oversimplification of the 
nuclear interactions, this is not a bad approximation as long as the
low density regimes, $\rho _{b} \lesssim 10^{14}$ g/cm$^{3}$ is
concerned, which is the case in this study, and gives some insight
into what the collisional effects are like and how important they
could be. Since the scatterings are
suppressed due to the Fermi blocking in the high density regime, it is 
expected the scattering effects are most important in the low density
regime. In addition, the resulting broadening of the structure
functions of nucleon will affect the neutrino energy spectra which are
formed in this low density regime\cite{hr98}. 

\par 
Some remarks are necessary. The above approximation is not applicable
to the vector current contribution to the neutrino scattering rates, 
since it does not take into
account the conservation law, and, as a result, the long wave length limit
is not properly reproduced. In the case of the axial vector current,
the spin density is not conserved if one takes into account the nuclear
tensor force. What is more important, the nuclear axial vector current 
is nearly isovector. Thus unless the matter is
entirely composed of neutrons or protons, the axial vector component
of the nuclear weak current is not a conserved one as long as the
nuclear force is spin-dependent.

\subsection{approximations}

Here we summarize the formulation outlined above. The spin correlation 
functions are given by the dressed Green function as
\begin{equation}
\label{eq:r2a}
R_{2}(k) \ = \ h_{A}^{2} \, 2 \, \int \frac{d^{4}p}{(2 \pi) ^{4}}
\ G_{N}^{>}(p + k) \  G_{N}^{<}(p) \quad ,
\end{equation}
where the nucleon Green functions are defined as
\begin{eqnarray}
i \, G_{N}^{>}(p) \ & \ = \ & \  \ \rho(p) \ [\, 1 \, - \, f_{N}(p) \, ] \\
i \, G_{N}^{<}(p) \ & \ = \ & \ - \, \rho(p) \ f_{N}(p) \qquad .
\end{eqnarray}
$f_{N}(p)$ is the Fermi-Dirac distribution of nucleon. 
$\rho(p)$ is the spectral function of nucleon given by 
the self-energy $\Sigma(p)$ as
\begin{equation}
\label{eq:rho}
\rho \, (p^{0}, \bbox{p}) \ = \ \frac{Im \Sigma(p)}
{\, \left [p^{0} - \displaystyle{\frac{\bbox{p}^{2}}{2M_{N}}} 
- Re \Sigma(p)\, \right ]^{2} 
\ + \ \left [\, \displaystyle{\frac{Im \Sigma(p)}{2}}\, \right ]^{2}} \qquad .
\end{equation}
$M_{N}$ is the nucleon mass. The quasi-particle approximation, which
is usually assumed, is obtained as a limit $Im \Sigma(p) \rightarrow
0$. In the Hartree or Hartree-Fock approximation, the self-energy is
real and the nucleon can be treated as a quasi-particle. Beyond those
approximations, the self-energy has an imaginary part in general. It is 
obvious that the imaginary part of the self-energy is a width of the
spectral function of the transfer energy $p^{0}$. Hence the spectral
function becomes a $\delta $ function in the quasi-particle
approximation. 

In this paper, following Alm et al.~\cite{ar96}, we evaluate the
imaginary part of the nucleon self-energy from the two-particle 
$T$-matrix which is given as a solution of the Bethe-Salpeter equation in the ladder
approximation,
\begin{eqnarray}
T(p_{1}, p_{2}, p_{1'}, p_{2'}, z) & \ = \ & V(p_{1}, p_{2}, p_{1'}, p_{2'}) 
\nonumber \\
\label{eq:bs}
& \ + \  &
\int \! \frac{d^{3}p_{3}}{(2 \pi )^{3}} \frac{d^{3}p_{4}}{(2 \pi )^{3}} 
\frac{d^{3}p_{5}}{(2 \pi )^{3}} \frac{d^{3}p_{6}}{(2 \pi )^{3}}
\ V(p_{1}, p_{2}, p_{3}, p_{4}) 
\ G_{2}^{0}(p_{3}, p_{4}, p_{5}, p_{6}, z)
\ T(p_{5}, p_{6}, p_{1'}, p_{2'}, z) \qquad .
\end{eqnarray}
Here $G_{2}^{0}$ is defined as the product of two single-particle
Green functions and calculated in the quasi-particle approximation as
\begin{equation}
G_{2}^{0}(p_{1}, p_{2}, p_{1}', p_{2}', z) \ = \ 
\frac{1 - f_{N}(p_{1}) - f_{N}(p_{2})}{z - E(p_{1}) - E(p_{2})} 
\ \delta ^{3}(p_{1} - p_{1}') \, \delta ^{3}(p_{2} - p_{2}')\quad .
\end{equation}
$E(p)$ is the on-shell energy determined from the real part of the
denominator of Eq.~(\ref{eq:rho}) and approximated as explained later.
Eq.~(\ref{eq:bs}) is solved for the separable Yamaguchi potential,
$V_{\alpha }(p, p') = w_{\alpha }(p) \lambda _{\alpha } 
w_{\alpha }(p')$, with $w_{\alpha }(p) = 
\frac{\lambda _{\alpha }}{p^{2} + \gamma ^{2}}$, where the coupling 
constant and the effective range are $\lambda _{\alpha } =
-12.3178$ (MeV fm$^{-1}$)$^{1/2}$ for $\alpha = {^{1} \! S_{0}}$ and 
$\lambda _{\alpha } = -14.6988$ (MeV fm$^{-1}$)$^{1/2}$ for 
$\alpha = {^{3} \! S_{1}}$ and 
$\gamma = 1.4488$ fm$^{-1}$, respectively. Then the imaginary part of
the self-energy is given by
\begin{equation}
\label{eq:imsig}
Im \Sigma (p, \omega + i \varepsilon ) \ = \ 
\int \! \frac{d^{3}p'}{(2 \pi )^{3}} \ \left [ \,
f_{N}(E(p')) \ + \ g_{B} (E(p') + \omega ) \, \right ] \ 
Im T_{ex}(p, p', p, p', E(p') + \omega + i \varepsilon ) \quad .
\end{equation} 
Here again the quasi-particle approximation is used for a
single-particle Green function. $g_{B}(E) = 
\{ \exp [(E - 2 \mu ) / T] - 1 \}^{-1}$ is a Bose-Einstein
distribution function with the chemical potential of the nucleon
pair. The subscript $ex$ denotes the inclusion of the 
exchange term. The real part of the self-energy is also derived from
the $T$-matrix in a similar way. Since it depends on the on-shell
energy which is in turn determined by the real part of the
self-energy, iterations are required to obtain the self-consistent
value of the on-shell energy. For simplification we abandon this
consistency in this paper and set the on-shell energy as $E(p) = p^{2} / 2
M_{N}^{*} + U$, where the effective mass $M_{N}^{*}$ and the potential $U$ 
are taken from the relativistic mean field theory\cite{st94a,st94b}. 
Using this on-shell 
energy and the corresponding imaginary part of the self-energy
calculated by Eq.~(\ref{eq:imsig}), we approximate Eq.~(\ref{eq:rho}) as $\rho(p) = 
Im \Sigma(p) / \{[\, p^{0} - E(p)\, ]^{2} 
+ [\, Im \Sigma(p) / 2 \, ]^{2} \}$. Finally we evaluate
Eq.~(\ref{eq:r2a}) with this approximate spectral function of nucleon.

\section{results}

\begin{figure}[thbp]
\vspace*{-2.5cm}
\hspace{4cm}
\epsfig{file=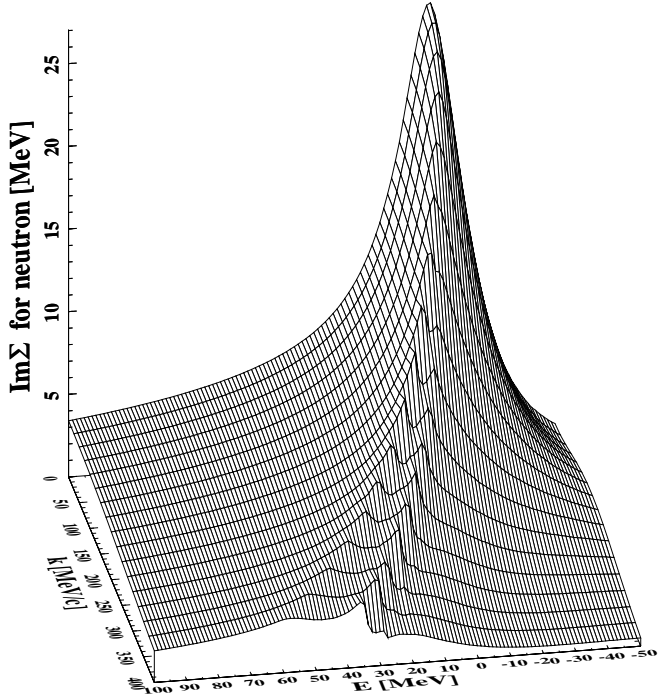,width=.65\textwidth}
\begin{center}
\caption{\label{fig3} The imaginary part of the self-energy for neutron as a
function of the transfer energy $E$ and the absolute value of the
transfer three momentum. 
The baryonic density is $\rho _{b} = 3 \times 10^{13}$g/cm$^{3}$,
the temperature is $T = 10$MeV and the electron fraction is $Y_{e} = 0.4$.}
\end{center}
\end{figure}

\begin{figure}[thbp]
\vspace*{-2.5cm}
\epsfig{file=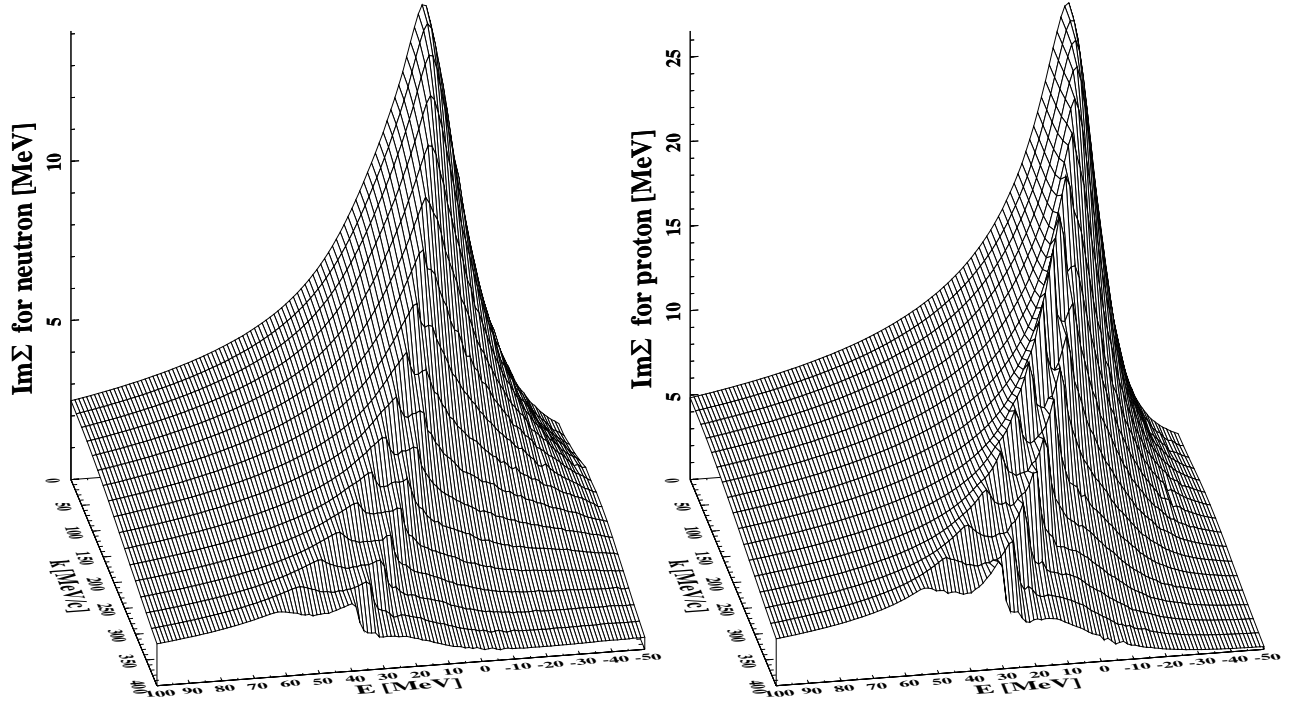,width=1.25\textwidth}
\begin{center}
\caption{\label{fig4} The imaginary part of the self-energy for
neutron (left) and proton (right) as a
function of the transfer energy $E$ and the absolute value of the
transfer three momentum. 
The baryonic density is $\rho _{b} = 3 \times 10^{13}$g/cm$^{3}$,
the temperature is $T = 10$MeV and the electron fraction is $Y_{e} = 0.1$.}
\end{center}
\end{figure}

\begin{figure}[thbp]
\vspace*{-2.0cm}
\hspace{4cm}
\epsfig{file=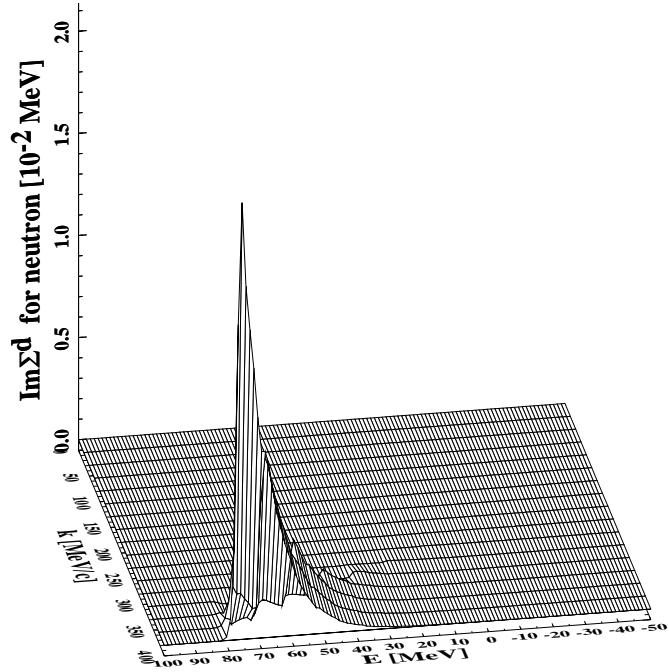,width=0.65\textwidth}
\begin{center}
\caption{\label{fig5} The deuteron contribution to the imaginary part 
of the self-energy for neutron as a
function of the transfer energy $E$ and the absolute value of the
transfer three momentum. 
The baryonic density is $\rho _{b} = 10^{13}$g/cm$^{3}$,
the temperature is $T = 10$MeV and the electron fraction is $Y_{e} = 0.4$.}
\end{center}
\end{figure}

In Fig.~\ref{fig3}, we show the imaginary part of the neutron 
self-energy as a function of the transfer momentum and energy for the
density $\rho _{b} = 3 \times 10^{13}$g/cm$^{3}$ and temperature $T =
10$MeV and electron fraction $Y_{e} = 0.4$. The on-shell momenta
roughly correspond to the ridge of the hill in this figure. The
imaginary part of the self-energy is essentially a scattering rate of 
a neutron with another nucleon. It is evident from the figure that 
this is energy- and momentum-dependent and becomes as large as the 
average kinetic energy of the nucleon $\sim 3 T$ for small momentum
transfers. $Im \Sigma$ decreases as
one goes to higher energies along the on-shell, since the scattering
rate becomes smaller for higher energy incident nucleons. The
self-energy of proton is almost the same as that of neutron in this case.
For more asymmetric matter with smaller $Y_{e}$, $Im \Sigma _{p}$
becomes larger while $Im \Sigma _{n}$ gets smaller as seen in
Fig.~\ref{fig4}. This is simply because the scattering rate for the
neutron -- proton pair is larger than that for the neutron -- neutron or
the proton -- proton pair. The greater degeneracy for neutron also
contributes to the reduction of the neutron self-energy. The density-, 
temperature- and $Y_{e}$-dependences of  $Im \Sigma $ can be understood
essentially in the same way. Since we
solved the Bethe-Salpeter equation, the contribution from the deuteron 
pole, if any, is automatically taken into account. In fact, the parameter
values of the Yamaguchi potential used in this paper are chosen 
so that the deuteron binding energy is reproduced in the vacuum. For the above
cases, we confirmed no bound state can exist due to the medium
effect as shown by Alm et al.\cite{ar96} For the lower density 
$\rho _{b} = 10^{13}$g/cm$^{3}$, however, 
the deuteron pole does exist and give tiny contributions to the
imaginary part of the self-energy as shown in Fig.~\ref{fig5}, where
only the contribution from the deuteron pole was shown. The
existence of the bound state is, thus,  not relevant for the density,
temperature and $Y_{e}$ of our interest.  

\begin{figure}[thbp]
\vspace*{-1.5cm}
\hspace{4cm}
\epsfig{file=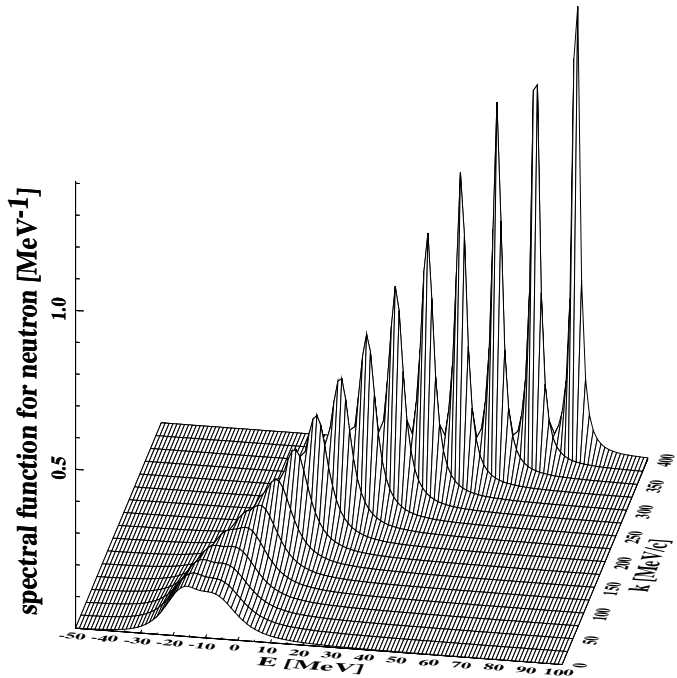,width=0.65\textwidth}
\begin{center}
\caption{\label{fig6} The spectral function for neutron as a
function of the transfer energy $E$ for various absolute values of the
transfer three momentum. 
The baryonic density is $\rho _{b} = 3 \times 10^{13}$g/cm$^{3}$,
the temperature is $T = 10$MeV and the electron fraction is $Y_{e} = 0.4$.}
\end{center}
\end{figure}

The spectral function is shown in Fig.~\ref{fig6} as a function of the 
transfer energy for different transfer momenta The on-shell
momentum is located again at the peak of the spectrum. As mentioned
above, this function would be a $\delta $-function of the transfer
energy if there were no interactions between nucleons. With collisions 
of two nucleons the spectral function is broadened, as expected, and 
the width of the spectral function is roughly the scattering rate 
for the given transfer momentum. It is seen that the width of the
spectral function or the scattering rate of nucleon is dependent on the 
transfer momentum and the spectral function becomes more
narrow-peaked as the transfer momentum gets greater, reflecting the
decrease of scattering rates. It should be again emphasized that the
width of the spectral function is of the same order for the low
momentum transfer as the average kinetic energy of nucleon. Hence it
is expected that the kinetics of neutrino -- nucleon scattering will
be also affected. The density-, temperature- and $Y_{e}$-dependences
of the spectral function can be inferred from that of the
imaginary part of the nucleon self-energy. As the density increases,
the scattering rate of nucleon with another one is increased, thus
leading to larger widths of the spectral function. Note that
the greater degeneracy of nucleon counteracts via the Fermi blocking 
to reduce the scattering rate. The scattering rate is determined by
this competition. As the temperature increases, the blocking effect is 
diminished. This enhances the scattering rate. However, the larger
kinetic energy of nucleon lowers the scattering rates. Again the width 
is determined by the balance of these factors. As the matter becomes
neutron rich, the width of the spectral function for neutron is
decreased partly because the scattering rate for neutron -- neutron
pair is smaller than that for neutron -- proton and partly because the 
neutron becomes more degenerate and the Fermi blocking reduces the
scattering rate more effectively.

\begin{figure}[thbp]
\vspace*{-1.0cm}
\hspace{3cm}
\epsfig{file=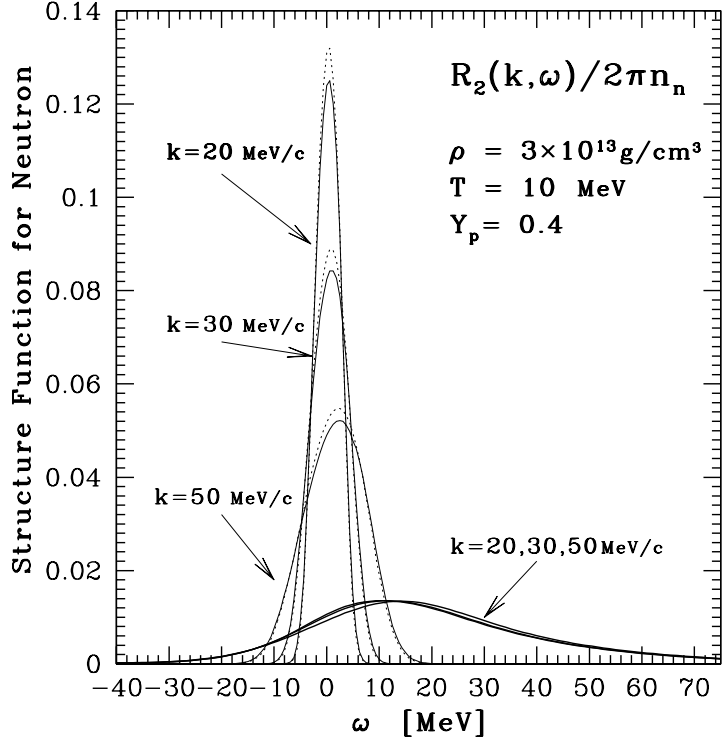,width=0.60\textwidth}
\begin{center}
\caption{\label{fig7} The dynamical structure function 
$R_{2}(k, \omega)$ for neutron as a
function of the transfer energy $\omega $ for various absolute values of the
transfer three momentum, $k$. 
The baryonic density is $\rho _{b} = 3 \times 10^{13}$g/cm$^{3}$,
the temperature is $T = 10$MeV and the electron fraction is $Y_{e} =
0.4$. The thin dotted lines represent 
the structure functions for the free neutron, while the thin solid lines 
present the case for which only the effective mass of nucleon was
taken into account in the spectral function, Eq.~(\protect\ref{eq:rho}).
The thick solid lines show the results with the collisional effects
taken into account for $k = 20, 30, 50$MeV/c from left to right.}
\end{center}
\end{figure}

Using the spectral functions obtained above, we calculated the nucleon 
structure functions with Eq.~(\ref{eq:r2a}). The typical result is
shown in Fig.~\ref{fig7} for $R_{2}(k, \omega )$ as a function of the
transfer energy $\omega $ for three different transfer momenta $k$. In 
this figure, three cases are compared. The thin dotted lines represent 
the structure functions for the free neutron, while the thin solid lines 
present the case for which only the effective mass of nucleon was
taken into account in the spectral function, Eq.~(\ref{eq:rho}). The
smaller effective mass lowers the peak of the spectral function as
pointed out by some authors\cite{re98b,re98c}, 
although this effect is small for this low 
density ($M_{N}^{*} = 890 MeV$). It also tends to broaden the
structure functions, since the recoil of the nucleon becomes larger
as the effective mass of nucleon gets smaller. Again this is
negligible in the low density regime considered here. On the other
hand, the structure 
function is considerably broadened when the collisions among nucleons
were taken into account, as shown with the thick solid lines in 
Fig.~\ref{fig7}. The width is essentially given by the scattering rate 
of nucleons and taken over from the corresponding spectral function. The widths are
not different for this low momentum transfers, which is qualitatively
different from the results for the free nucleons and those for the
quasi-particle approximation. As the density increases, the width
becomes larger in general. As the temperature decreases, the part of
the negative energy transfer is reduced, since the extraction of energy
from the medium becomes more and more difficult as the nucleons become 
more degenerate. This is nothing but the detailed balance relation 
$R(k) \ = \ e ^{\beta \omega} \ R(-k)$ which is
automatically satisfied in the formulation used in this paper. The
width of the structure function becomes smaller for neutron and greater
for proton as $Y_{e}$ is decreased, as expected from the above
discussions for the spectral function. 

The total scattering rate of the neutrino with the incident energy 
$E_{\nu }^{in}$ is given by the integration of the structure function 
with respect to the transferred momentum $k$ 
and energy $\omega $: 
\begin{eqnarray}
\label{eq:toteq}
R^{tot}(E_{\nu }^{in}) & = & \int \! \frac{d^{3} q_{\nu }^{out}}
{(2 \pi )^{3}} \, \frac{1}{2 E_{\nu }^{in} \, 2 E_{\nu }^{out}} \  
R(E_{\nu }^{in}, \, E_{\nu }^{out}, \, \cos \theta ) 
\  \left [ \, 1 \, - \, f_{\nu }(E_{\nu }^{out}) \, \right ]\nonumber \\
& = & \frac{1}{(2 \pi )^{3}} \, \int ^{\infty}_{0} \! 
2 \pi k \, dk \, \int _{-k}^{\omega ^{max}} \! \! \! d \omega \ 
\frac{E_{\nu }^{in} - \omega }{E_{\nu }^{in}} 
\frac{1}{2 E_{\nu }^{in} \, 2 E_{\nu }^{out}} \ 
R(E_{\nu }^{in}, \, E_{\nu }^{out}, \, \cos \theta ) 
\  \left [ \, 1 \, - \, f_{\nu }(E_{\nu }^{out})
\, \right ] ,
\end{eqnarray}
with $\omega ^{max} = min(k, \, 2E_{\nu }^{in} - k)$. The scattering
angle $\theta $ and the energy of the outgoing neutrino 
$E_{\nu }^{out}$ are functions of $E_{\nu }^{in}$ and $k$. 
The collisions of nucleons reduce in two ways the above scattering rates from
the quasi-particle approximation. Not only the amplitude of the
structure function is lowered, but the broadening of the structure
function tends to put some fraction of the structure function outside
the integration region. Ignoring the Fermi blocking of the outgoing
neutrino in Eq.~(\ref{eq:toteq}), we integrated the axial vector part 
$R_{2}^{tot}(E_{\nu }^{in})$ for $E_{\nu }^{in} = 3 T$ and compared the
result with the collision taken into account and the result for the
free nucleon. For the model shown in Fig.~\ref{fig7}, for example, we
found that the scattering rate is suppressed by the factor of 0.41 for
neutron and 0.38 for proton, respectively. Note that the contribution
of the effective mass is only a few percent in this case. For 
$\rho _{b} = 10^{13}$g/cm$^{3}$, $T = 10$MeV and $Y_{e} = 0.4$, we still found
the reduction of $\sim 0.6$. 

The larger width of the structure function implies that the neutrino
can exchange energy with nucleons more efficiently. In the supernova
simulations performed so far, the neutrino -- nucleon scattering was 
commonly approximated as the iso-energetic scattering, which means
that this process does not contribute to the thermalization of the
neutrino spectra. The above results stand by the claim by Raffelt and
his company\cite{rs95,rs96,rs98,hr98} that this may not be the case. 
In fact, the mean square
root of energy transfer increased from $\sim 5$MeV to $\sim 15$MeV for 
the model of Fig.~\ref{fig7}. Although this is still not a large
value, it should be born in my mind that this process has about ten
times larger cross section than the scattering with electrons. Hence
it is expected the energy spectra of neutrinos, particularly muon- and
tau-neutrinos, will be affected by this effect. Thus we need to somehow implement
this effect in a numerical code and to study quantitatively the
difference it will make in the realistic settings.

\section{conclusion}
 
We studied in this paper the possible effects of collisions among nucleons on
the reaction rates of the neutrino -- nucleon scatterings. By solving
the Bethe-Salpeter equation for two nucleons, we estimated the
imaginary parts of the nucleon self-energy, which in turn give the
spectral function of the nucleon in medium. Using this spectral
function, we calculated the structure functions of nucleon, which are
directly related to the neutrino -- nucleon scattering rates. In the limit of no
collision, the spectral function reduces to the delta function of the
transfer energy and the structure function gives the reaction rates
which have been commonly used in the literature thus far. It was shown 
that the collisional broadening of the structure functions might be the 
main factor in the low density regime considered in this paper to
modify not only the neutrino reaction rates with the nucleon but also
the neutrino spectra.

It is obvious that the further study is needed on this issue. Although 
we used Yamaguchi potential to facilitate calculations, this is
clearly an oversimplification of the dynamics of two nucleons. The
self-consistency was also sacrificed for simplicity. Thus our results 
should be regarded as of qualitative nature. More importantly, 
we have to assess the corrections of the terms we ignored in estimating the 
the structure functions, Eq.~(\ref{eq:sik}). 
The in-medium vertex corrections should be also taken into account and
are inferred from the Fermi liquid theory, for example\cite{vs87,sd99}. 
Though we neglected in this paper these corrections in order to make
clear the effect of the width of the spectral function of nucleon, 
they could be as important and should be studied on the
same basis in the future work. It should be also mentioned that 
the formulation in this paper is not conserving\cite{yt99,bk61,by62} and 
cannot be applied to the vector current, whose
long wave length limit is dictated by the baryon number
conservation. The accomodation of this limit should be another subject 
of the future work.

At present the neutrino transport is thought to play a crucial role in the
collapse-driven supernovae. It is, thus, important to study the
microphysics involved and, at the same time, to investigate its
consequences to the supernova simulations, which we are planning to do 
with the numerical code recently developed by us to solve the Boltzmann equations
for neutrinos\cite{yjs99}.

\acknowledgments

I gratefully acknowledge critical discussions
with H.-Th. Janka, G. Raffelt and H. Toki. I would highly appreciate various
support by E. M\"{u}ller during the stay in Max Planck Institut
f\"{u}r Astrophysik. This work is partially supported by the Japanese 
Society for the Promotion of Science (JSPS), Postdoctoral Fellowships 
for Research Abroad and by the 
Grants-in-Aid for the Center-of-Excellence (COE) Research of
the Ministry of Education, Science, Sports and Culture of Japan to
RESCEU (No.07CE2002).

% now the references. delete or change fake bibitem. delete next three
%   lines and directly read in your .bbl file if you use bibtex.

% \begin{figure}
% \caption{}
% \label{}
% \end{figure}

% \begin{table}
% \caption{}
% \label{}
% \end{tabular}
% \end{table}


\begin{references}
\bibitem{bz34a} W. Baade and F. Zwicky, Phys. Rev. {\bf 45}, 138 (1934)
\bibitem{bz34b} W. Baade and F. Zwicky, Phys. Rev. {\bf 46}, 76 (1934)
\bibitem{imb87} R. M. Bionta et al., Phys. Rev. Lett. {\bf 58}, 1494 (1987)
\bibitem{kam87} K. Hirata et al., Phys. Rev. Lett. {\bf 58}, 1490 (1987)
\bibitem{wil82} J. R. Wilson, Proc. Univ. Illinois Meeting on
                Numerical Astrophysics (1982)
\bibitem{bw85}  H. A. Bethe and J. R. Wilson, Astrophys. J. {\bf 295},
                14 (1985) 
\bibitem{bg93}  A. Burrow and J. Goshy, Astrophys. J. 
                {\bf 416}, 75 (1993)
\bibitem{jm93}  H.-Th. Janka and E. M\"{u}ller, Frontiers of Neutrino
                Astrophysics, Proc. of the International Symposium on Neutrino
                Astrophysics, Takayama/Kamioka, Japan, Oct. 19.-22., 1992, 
                edited by Y. Suzuki and K. Nakamura 
               (Universal Academy Press, Tokyo, 1993), p203
\bibitem{jm96}  H.-Th. Janka and E. M\"{u}ller, Astron. Astrophys. 
               {\bf 306}, 167 (1996)
\bibitem{he94}  M. Herant, W. Benz, J. Hix, C. Freyer and S. A. Colgate,
                Astrophys. J. {\bf 435}, 339 (1994)
\bibitem{bu95}  A. Burrows, J. Hayes and B. A. Fryxell, Astrophys. J. 
               {\bf 450}, 830 (1995)
\bibitem{sh94}  T. Shimizu, S. Yamada and K. Sato, Astrophys. J. Lett.
               {\bf 432}, L119 (1994)
\bibitem{mmbg98}  O. E. B. Messer, A. Mezzacappa, S. W. Bruenn and
                M. W. Guidry, Astrophys. J. {\bf 507}, 353 (1998)
\bibitem{yjs99} S. Yamada, H.-Th. Janka and H. Suzuki, Astron. Astrophys. 
               {\bf 344}, 533 (1999)
\bibitem{bu99}  A. Burrows, T. Young, P. A. Pinto, R. Eastman and
                T. Thompson, submitted to Astrophys. J. (1999)
\bibitem{br85}  S. W. Bruenn, Astrophys. J. Suppl. {\bf 58}, 771 (1985) 
\bibitem{mb93a} A. Mezzacappa and S. W. Bruenn, Astrophys. J. 
               {\bf 405}, 637 (1993) 
\bibitem{mb93b} A. Mezzacappa and S. W. Bruenn, Astrophys. J. 
               {\bf 405}, 669 (1993) 
\bibitem{mb93c} A. Mezzacappa and S. W. Bruenn, Astrophys. J.
               {\bf 410}, 740 (1993) 
\bibitem{iw82}  N. Iwamoto and C. J. Pethick, Phys. Rev. D {\bf 25}, 313 (1982)
\bibitem{sa89}  R. F. Sawyer, Phys. Rev. C {\bf 40}, 865 (1989)
\bibitem{hw91}  C. J. Horowitz and K. Wehrberger, 
                Phys. Lett. B {\bf 266}, 236 (1991)
\bibitem{sw95}  R. F. Sawyer, Phys. Rev. Lett. {\bf 75}, 2260 (1995)
\bibitem{rs95}  G, Raffelt and D. Seckel, PRD. {\bf 52}, 1780 (1995)
\bibitem{rs96}  G, Raffelt, D. Seckel and G. Sigl, PRD. {\bf 54}, 2784 (1996)
\bibitem{rs98}  G, Raffelt and G. Sigl, sumitted PRD. (1998)
\bibitem{bs98}  A. Burrow and R. F. Sawyer, PRC. {\bf 58}, 554 (1998)
\bibitem{bs99}  A. Burrow and R. F. Sawyer, PRC. {\bf 59}, 510 (1999)
\bibitem{re98b} S. Reddy, M. Prakash, J. M. Lattimer and J. A. Pons,
                preprint Astro-ph/9811294 (1998)
\bibitem{re98c} S. Reddy, M. Prakash and J. M. Lattimer, Proc. Second
                Oak Ridge Symposium on Atomic and Nuclear Astrophysics, 1998 
\bibitem{yt99}  S. Yamada and H. Toki, submittded to Phys. Rev. C (1999)
\bibitem{lp53a} L. D. Landau and I. Pomeranchuk,
                Dokl. Akad. Nauk. SSSR {\bf 92}, 535 (1953)
\bibitem{lp53b} L. D. Landau and I. Pomeranchuk,
                Dokl. Akad. Nauk. SSSR {\bf 92}, 735 (1953)
\bibitem{mg56}  A. B. Migdal, Phys. Rev. {\bf 103}, 1811 (1956)
\bibitem{hr98}  S. Hannestad and G. Raffelt, Astrophys. J. 
               {\bf 507}, 339 (1998)
\bibitem{srs90} M. Schmidt, G. R\"{o}pke and H. Schulz, Ann. Phys.
               {\bf 202}, 57 (1990)
\bibitem{ar96}  T. Alm, A. Schnell, N. H. Kwong and H. S. K\"{o}hler,
                Phys. Rev. C {\bf 53}, 2181 (1996)
\bibitem{rs99}  G. R\"{o}pke and A. Schnell, Prog. Part. Nucl. Phys. 
               {\bf 42}, 53 (1999) 
\bibitem{kv95}  J. Knoll and D. N. Voskresensky, Phys. Lett. B 
               {\bf 351}, 43 (1995)
\bibitem{ym54}  Y. Yamaguchi, Phys. Rev. {\bf 95}, 1628 (1954)
\bibitem{st94a} K. Sumiyoshi and H. Toki, Astropphys. J. {\bf 422}, 700 (1994)
\bibitem{st94b} Y. Sugahara and H. Toki,  Nucl. Phys. A. {\bf 579}, 557 (1994)
\bibitem{vs87}  D. N. Voskresensky and A. V. Senatorov,
                Sov. J. Nucl. Phys. {\bf 45}, 411 (1987)
\bibitem{sd99}  A. Sedrakian and A. Dieperink, preprint
                nucl-th/9905039 (1999)
\bibitem{bk61}  G. Baym and L. P. Kadanoff, Phys. Rev. {\bf 124}, 287 (1961)
\bibitem{by62}  G. Baym, Phys. Rev. {\bf 127}, 1391 (1962)
\end{references}
\end{document}